\documentclass[12pt]{article}
\usepackage{cite,psfig}

\newcommand{\newc}{\newcommand}
\newc{\gsim}{\lower.7ex\hbox{$\;\stackrel{\textstyle>}{\sim}\;$}}
\newc{\lsim}{\lower.7ex\hbox{$\;\stackrel{\textstyle<}{\sim}\;$}}
\newc{\gev}{\,{\rm GeV}}
\newc{\mev}{\,{\rm MeV}}
\newc{\ev}{\,{\rm eV}}
\newc{\kev}{\,{\rm keV}}
\newc{\tev}{\,{\rm TeV}}

\def\tr{\mathop{\rm tr}}

\newc{\mz}{M_Z}
\newc{\mpl}{M_*}
\newc{\mw}{m_{\rm weak}}
\def\beq{\begin{equation}}
\def\eeq{\end{equation}}
\def\bea{\begin{eqnarray}}
\def\eea{\end{eqnarray}}
\newc{\ie}{{\it i.e.}}          \newc{\etal}{{\it et al.}}
\newc{\eg}{{\it e.g.}}          \newc{\etc}{{\it etc.}}
\newc{\cf}{{\it c.f.}}
\def\bar#1{\overline{#1}}

\def\inv{^{\raise.15ex\hbox{${\scriptscriptstyle -}$}\kern-.05em 1}}
\def\lbar{{\lower.35ex\hbox{$\mathchar'26$}\mkern-10mu\lambda}} 

\def\to{\rightarrow}

\let\p=\partial
\let\<=\langle
\let\>=\rangle

\let\+=\uparrow
\let\al=\alpha

\let\de=\delta

\let\la=\lambda

\let\si=\sigma
\let\Si=\Sigma
\let\th=\theta

\renewcommand{\epsilon}{\varepsilon}
\renewcommand{\phi}{\varphi}

\begin{document}
\thispagestyle{empty}
\vspace*{.5cm}
\noindent
\hspace*{\fill}{CERN-TH/2001-172}\\
\hspace*{\fill}{LBNL-48449}

\vspace*{2.0cm}

\begin{center}
{\Large\bf The Structure of GUT Breaking by Orbifolding}
\\[2.5cm]
{\large Arthur Hebecker$^*$ and John March-Russell$^{*,\dagger}$
}\\[.5cm]
{\it $^*$Theory Division, CERN, CH-1211 Geneva 23, Switzerland}
\\[.2cm]
{\it $^\dagger$Theory Group, LBNL, University of California,
Berkeley, CA 94720, USA}
\\[.2cm]
(July 4, 2001)
\\[1.1cm]

{\bf Abstract}\end{center}
\noindent
Recently, an attractive model of GUT breaking has been proposed in which 
a 5 dimensional supersymmetric SU(5) gauge theory on an $S^1/(Z_2\times 
Z_2')$ orbifold is broken down to the 4d MSSM by SU(5)-violating boundary 
conditions. Motivated by this construction and several related realistic
models, we investigate the general structure of orbifolds in the effective
field theory context, and of this orbifold symmetry breaking mechanism
in particular. An analysis of the group theoretic structure of orbifold
breaking is performed.  This depends upon the existence of appropriate
inner and outer automorphisms of the Lie algebra, and we show
that a reduction of the rank of the GUT group is possible.  Some
aspects of larger GUT theories based on SO(10) and E$_6$ are discussed.
We explore the possibilities of defining the theory 
directly on a space with boundaries and breaking the gauge symmetry by 
more general consistently chosen boundary conditions for the fields.
Furthermore, we derive the relation of orbifold breaking with the
familiar mechanism of Wilson line breaking, finding a one-to-one
correspondence, both conceptually and technically.  Finally, we analyse
the consistency of orbifold models in the effective field theory
context, emphasizing the necessity for self-adjoint extensions of
the Hamiltonian and other conserved operators, and especially the
highly restrictive anomaly cancellation conditions that apply if
the bulk theory lives in more than 5 dimensions. 
\newpage

\section{Introduction}
The paradigm of grand unification has dominated our thinking about physics 
at very high energies since the pioneering work of Georgi and 
Glashow~\cite{GG} (also~\cite{PS}). The success of gauge coupling 
unification~\cite{GQW} in supersymmetric extensions of these grand unified 
theories (GUTs)~\cite{DRW} has further supported this idea~\cite{unif}. 
However, the GUT concept has well-known problems, such as the Higgs 
structure at the high scale (especially doublet-triplet splitting), 
the issue of too fast proton decay, and the mismatch of the GUT scale with 
the naive scale of unification with gravity.

A new possibility for embedding of the standard model (SM) into a form of
GUT has been suggested by Kawamura~\cite{kaw1,kaw2,kaw3} and further
extended by Altarelli and Feruglio~\cite{AF} and by Hall and
Nomura~\cite{HN} (see also~\cite{kk,HMR}).  The basic idea is
that the GUT gauge symmetry is realized in 5 or more space-time
dimensions and only broken down to the SM by utilizing GUT-symmetry
violating boundary conditions on a singular `orbifold' compactification.
Given the success of supersymmetric gauge-coupling unification, the most 
attractive models include both supersymmetry and (at least) SU(5) gauge 
symmetry in 5 dimensions.  In these models, both the GUT group and 5d 
supersymmetry (corresponding to N=2 SUSY in 4d) are broken down to a N=1 
supersymmetric model with SM gauge group by compactification on $S^1/(Z_2 
\times Z_2')$ (related ideas were employed for electroweak and low-energy 
SUSY breaking; see, e.g.,~\cite{BHN}). One of the most attractive features 
of this construction is the solution of the doublet-triplet splitting 
problem by boundary conditions for the Higgs which are closely linked to the 
breaking pattern of SU(5). More recently, it has been observed that an even
simpler way of removing the triplet Higgs is provided by localizing the 
Higgs field at the SU(5)-breaking brane~\cite{HMR}. A further motivation 
for the above orbifold GUTs follows from string theory, which requires 
both additional dimensions as well as branes located at orbifold fixed 
points. Thus, taking the phenomenological success of traditional gauge 
coupling unification seriously, the energy range between the GUT scale and 
the string (or Planck) scale is the natural domain for higher-dimensional 
field theories. In the following, we will take an effective-field-theory 
viewpoint of orbifold branes and impose all necessary constraints of a 
consistent low-energy theory without requiring an explicit string-theory 
realization.

In this paper, we analyse the generic structure of orbifold breaking of gauge 
symmetries, illustrating our results with phenomenologically important 
examples.\footnote{
After 
this paper has been completed, a related investigation of structural 
issues in orbifolding appeared~\cite{bhn1}. Although there is some overlap
with our discussion, most results are complementary since~\cite{bhn1} 
emphasizes SUSY breaking and TeV scale models, while we focus on gauge 
symmetry breaking with applications mainly at the GUT scale.
}
Section~2.1 introduces abstractly the field-theoretic orbifolding 
procedure, which is based on a discrete symmetry group acting in physical 
space and in field space. The popular model of SU(5) breaking on a 
$S^1/(Z_2\times Z_2')$ orbifold is described in Sect.~2.2. The group 
theoretic structure of orbifold breaking of gauge symmetries is analysed 
in detail in Sect.~3. Different breaking patterns emerge if the discrete 
orbifolding symmetry is realized in field space as an inner (see Sect.~3.1) 
or outer (see Sect.~3.2) automorphism of the Lie algebra. A complete 
classification can be given in the physically most interesting case of a
$Z_2$ symmetry. In Sect.~4, we advocate an alternative and much more 
general approach, which starts from a theory defined on a space with 
boundaries and breaks the gauge symmetry by consistently chosen 
boundary conditions for the gauge potential. Section~4.1, which 
discusses Dirichlet and Neumann boundary conditions, is followed by a 
critical assessment of more restrictive boundary conditions in Sect.~4.2 
and by a discussion of the dynamical realization of boundary 
conditions by expectation values of boundary fields in Sect.~4.3.
For example, the physically important breaking of SO(10) to SU(5), 
inaccessible to the $Z_2$ orbifolding procedure, is straightforwardly 
realized by a boundary scalar in the {\bf 16} of SO(10). 
The close relation of orbifold breaking and Wilson line breaking of gauge 
symmetries, which is apparent at a conceptual level, is treated in 
technical detail in Sect.~5. In the orbifolding case, the background gauge 
field responsible for the non-trivial Wilson loop is restricted to the 
singularity, where it enforces the typical orbifolding boundary conditions 
for the gauge potential. The quantum field theoretic consistency 
of orbifold models is discussed in Sect.~6. Section 6.1, dealing with the 
elementary constraints of unitarity and self-adjointness, is followed by
Sect.~6.2, where anomaly cancellation is discussed and found to be highly 
restrictive in more than 5 dimensions. Our conclusions are given in 
Sect.~7.

\section{Orbifold Breaking}\label{orbb}

In this section we discuss more generally the idea of `orbifolding' a 
quantum field theory with gauge group $G$. We then describe how the 
particularly popular example of SU(5) breaking on $S^1/(Z_2\times 
Z_2')$ fits into this framework. 

\subsection{The meaning of orbifolding}\label{meaning}

Consider a (higher-dimensional) QFT with gauge group $G$ defined on a 
manifold $M=R^4\times C$ ($C$ has coordinates $y^i$, $i=1,\ldots,{\rm dim} 
(C)$). Suppose that both the manifold $C$ and the QFT possess a symmetry
under a discrete group $K$. The action of $K$ on the internal manifold 
$C$ is geometrical,
\beq
K: (x,y) \to (x,k[y]),
\label{eq:geom}
\eeq
where $k[y]$ is the image of the point $y$ under the operation
of $k\in K$, while the action of $K$ in field space is
\beq
K: \Phi_i \to R(k)_{ij}\Phi_j .
\label{eq:fields}
\eeq
Here $\Phi$ is a vector comprising all fields in the theory, and $R(k)$
is a, possibly reducible, matrix representation of $K$.

We can orbifold or `mod out' the theory by $K$ by declaring that only field 
configurations invariant under the actions Eqs.~(\ref{eq:geom}) and 
(\ref{eq:fields}) are physical. Alternatively, one can replace 
Eq.~(\ref{eq:fields}) by the trivial action (in which case $K$ acts purely 
geometrically), or one can replace Eq.~(\ref{eq:geom}) by the trivial 
action (in which case $K$ acts purely in field space). Let us discuss the 
physical meaning of these three possibilities in turn.

Modding out by just the geometrical action Eq.~(\ref{eq:geom}) means
that instead of working on the physical space $C$, we define the theory 
on $C/K$.  First let $K$ act freely.  Namely,
\beq
k[y] \neq y, \qquad \forall y\in C, \quad \forall k\neq 1\in K ,
\eeq
so that non-trivial elements of $K$ move all points of $C$.
Then the space $C/K$ is smooth and is again a manifold.
Note that we have not reduced the amount of gauge symmetry
of the theory; it is still $G$, but now defined on the smaller physical
space $R^4\times C/K$.
A relevant example of such a construction is to take $C$ to be the real
line which we then mod out by the equivalence $y\sim y+2\pi R$ generated
by the discrete translation group $K={\bf Z}$, leading to the smooth
space $R/{\bf Z} = S^1$ of radius $R$.

On the other hand when the action of $K$ has fixed points ($k[y]=y$
for some $y\in C$, $k\neq 1$), then $C/K$ is not smooth, having
singularities at the fixed points.  Such a space is known as an
orbifold.
A classic result of string theory is that string propagation on such 
singular manifolds is well-defined~\cite{orbifold}.  Here, following 
Refs.~\cite{kaw1,kaw2,kaw3,kk,AF,HN,HMR} (see also~\cite{fmt}), we wish to 
consider orbifolds in the effective field theory context, in which the 
consistency of propagation on such singular spaces has to be 
re-assessed~\cite{AHCG}. This we postpone to Section~\ref{consistency}.

The second possibility is that we mod out by just the action
Eq.~(\ref{eq:fields}) on field space.  In this case the geometrical
nature of the physical space is unchanged, it is still $R^4\times C$.
However the amount of gauge symmetry has been reduced to the centralizer
$N_K(G)$ of $K$ in $G$ ($K$ is now to be thought of as a subgroup of $G$)
\footnote{This is not precisely correct.  The representation $R(k)$ by which
$K$ acts on the space of fields may not be faithful, in which case
the surviving group is just the centralizer in $G$ of the faithfully
represented subgroup of $K$.}.
The reason for the breaking is that $K\subset G$ does
not commute with all of $G$ (we will study this in greater
detail in Section~\ref{inner}).  Recall that
the centralizer of an element $k$ in a group $G$ is defined as
\beq
N_k(G) \equiv \{ g\in G : gk=kg \}\,.
\eeq
Such modding out reduces the size of the gauge group everywhere in 
$R^4\times C$ to the smaller gauge group $N_K(G)\equiv H$.  This is just as 
if we had started with the smaller gauge theory based on $H$.\footnote{
Even this seemingly trivial form of `orbifolding' leads to interesting
information about the relation of the parent and daughter theories, \eg,
in the large-N limits. Specifically there exists an `inheritance
principle' relating the correlation
functions~\cite{bershadsky,AHDMR,strassler}.}
As a side remark, let us note that one can start directly from a 4d SU(5)
theory and reduce it to the SM by modding out just in field space. Using a
$Z_2$ symmetry along the lines of Kawamura's proposal (see Sect.~\ref{ex} 
below), one can dispose of the triplet Higgs at the same time. Of course, 
one has now no fundamental reason forbidding additional SU(5)-violating 
operators (which could spoil gauge coupling unification completely). 
However, since one is in the weakly coupled regime, there is also no 
apparent dynamical mechanism generating such dangerous terms.

The third possibility is that the equivalence involves a
simultaneous action on both coordinates and field space,
\beq
\Phi_i(x,y)\sim R(k)_{ij}\Phi_j(x,k^{-1}[y])\, .
\label{eq:orbifold}
\eeq
The theories of particular interest are ones where the action of
$K$ on $C$ is {\em not} free, so that the geometrical space
is an orbifold $C/K$ with singular points. The special feature of
this case is that away from these fixed points
the gauge symmetry remains $G$, but {\em at the fixed points}
it is reduced to a subgroup $H\subset G$.  This group is determined as
follows: At each $y$ consider the discrete subgroup $F_y\subset K$ of
elements $k\in K$ that leave $y$ fixed, $F_y\equiv\{ k\in K : k[y]=y\}$.
Then the unbroken gauge group at $y$ is the centralizer of $F_y$ in $G$,
\beq
H_y = \{ g\in G : gk=kg, \forall k\in F_y \} .
\eeq
This combines features of the first and second possibilities
in a particularly interesting way.

A recent and physically relevant example is provided by the SU(5)
GUT model of Refs.~\cite{kaw2,AF,HN,HMR} based on an
$S^1/(Z_2\times Z_2')$ orbifold.  Since we will be interested in asking
how this model can be generalized it is useful for us to summarize some
of its essential features, a task to which we now turn.

\subsection{An example: The $S^1/(Z_2\times Z'_2)$ model}\label{ex}

Consider a 5-dimensional factorized space-time comprising
a product of 4d Minkowski space $R^4$ (with coordinates $x^{\mu}$,
$\mu=0,\ldots,3$), and the orbifold $S^1/(Z_2\times Z_2')$, with coordinate
$y\equiv x^5$.  The circle $S^1$ has radius
$R$ where $1/R\sim M_{\rm GUT}$.
The orbifold $S^1/Z_2$ is obtained by modding out the theory
by a $Z_2$ transformation which imposes on fields which depend upon
the 5th coordinate the equivalence relation $y\sim -y$.
To obtain the orbifold $S^1/(Z_2\times Z_2')$ we further mod out
by $Z'_2$ which imposes the equivalence relation $y'\sim -y'$,
with $y'\equiv y+\pi R/2$.
With the basis of identifications
\beq
P:~~y\sim -y \qquad P':~~y' \sim -y' .
\label{eq:PPequiv}
\eeq
there are two inequivalent fixed 3-branes (or `orbifold planes')
located at $y=0$, and $y =\pi R/2\equiv\ell$, which we denote $O$ and $O'$
respectively.  It is consistent to work with the theory obtained by
truncating to the physically irreducible interval
$y\in [0,\ell]$ with the 3-branes at $y=0,\ell$ acting as
`end-of-the-world' branes~\footnote{
Note that the discrete group $Z_2\times Z_2'$, generated by $P$ and $P'$, 
can also be considered as being generated by $P$ and $PP'$. The fact that 
the generator $PP'$ is a translation (i.e., it acts freely) suggests a close 
relation to Wilson line breaking. This relation, which can also be discussed 
in terms of the original generators $P$ and $P'$, is explored in detail in 
Sect.~\ref{wilson}.}.

The action of the equivalences $P,P'$ on the fields of a quantum field
theory living on $R^4\times S^1/(Z_2\times Z_2')$ is not fully
specified by the action Eq.~(\ref{eq:PPequiv}) on the coordinates.  One
must also define the action within the space of fields.  To this end, let
$\Phi(x,y)$ be a vector comprising all bulk fields, then the action
of $P$ and $P'$ is given by $P: \Phi(x,y)\sim P_\Phi \Phi(x,-y)$ and
$P': \Phi(x,y')\sim  P'_\Phi \Phi(x,-y')$.
Here, $P_\Phi$ and $P'_\Phi$ are matrix representations
of the two $Z_2$ operator actions, with eigenvalues $\pm 1$.
Let us from now on work in this diagonal basis of fields, and
classify the fields by their $(P,P')$ eigenvalues $(\pm 1,\pm 1)$. Then
the fields $\Phi_{PP'}(x,y)$ have KK expansions which involve
$\cos(k y/ R)$ with $k=2n$ or $2n+1$, for $\Phi_{+P'}$ ($P'=+,-$
respectively), and $\sin(ky/R)$ with $k=2n+1$ or $2n+2$, for
$\Phi_{-P'}$ ($P'=+,-$ respectively).
From the 4d perspective the KK modes acquire a mass $k/R$,
so only the $\Phi_{++}$ possess a massless zero mode.  Moreover,
only $\Phi_{++}$ and $\Phi_{+-}$
have non-zero values at $y=0$, while only $\Phi_{++}$ and $\Phi_{-+}$
are non-vanishing at $y=\ell$.
The action of the identifications
$P,P'$ on the fields (namely the matrices $P_\Phi$ and $P'_\Phi$)
can utilize {\em all} of the symmetries of the
bulk theory.  Thus $P$ and $P'$ can involve gauge transformations,
discrete parity transformations, and in the supersymmetric
case, R-symmetry transformations.

To reproduce the good predictions of a minimal supersymmetric GUT,
one starts from a 5d SU(5) gauge theory with minimal SUSY in 5d (with 8 real
supercharges, corresponding to N=2 SUSY in 4d).  Thus, at minimum, the
bulk must have the 5d vector superfield, which in terms of 4d N=1 SUSY
language contains a vector supermultiplet $V$ with physical components
$A_\mu,\la$, and a chiral multiplet $\Si$ with components $\psi,\si$.
Both $V$ and $\Si$ transform in the adjoint representation of SU(5).

If the parity assignments, expressed in the fundamental representation
of SU(5), are chosen to be
$P={\rm diag}(+1,+1,+1,+1,+1)$, and $P'={\rm diag}(-1,-1,-1,+1,+1)$,
so that the equivalence under $P$ is $V^a(x,y)T^a \sim
V^a(x,-y)P T^a P^{-1}$, and similarly for $P'$, then SU(5) is broken
to SU(3)$\times$SU(2)$\times$U(1) on the $O'$ fixed-brane, but is
unbroken in the bulk and on $O$.
If for $\Si$ the same assignments are taken apart from an overall sign
for both $P$ and $P'$ equivalences, \eg, under $P'$,
$\Si^a(x,y')T^a  \sim - \Si^a(x,-y') P' T^a P'^{-1}$, then
these boundary conditions also break
4d N=2 SUSY to 4d N=1 SUSY on both the $O$ and $O'$ branes.
Only the $(+,+)$ fields possess massless zero modes, and at low energies
the gauge and gaugino content of the 4d N=1 MSSM is apparent.
{\begin{center}
\begin{tabular}{|c|c|c|}
\hline
$(P,P')$ & 4d superfield & 4d mass\\
\hline
$(+,+)$ &  $V^a$ & $2n/R$\\
$(+,-)$ &  $V^{\hat{a}}$ & $(2n+1)/R$ \\
$(-,+)$ &  $\Si^{\hat{a}}$ & $(2n+1)/R$\\
$(-,-)$ &  $\Si^{a}$ & $(2n+2)/R$ \\
\hline
\end{tabular}
\end{center}}
\vspace{3mm}
Table 1. Parity assignment and KK masses of fields in the 4d vector
and chiral adjoint supermultiplet. The index $a$ labels the
unbroken SU(3)$\times$SU(2)$\times$U(1) generators of SU(5),
while $\hat{a}$ labels the broken generators.

In summary, the general situation is that, if $K$ acts on the extra
dimensional manifold, $C$, non-freely and
its action in field space does not commute with $G$, then the resulting
theory has a smaller gauge symmetry $H\subset G$. The symmetry breaking is
localized on (3+1)-dimensional submanifolds, which correspond to the fixed
points (actually fixed 3-branes) in $R^4\times C$ under the action of $K$.

\section{The group-theoretic structure of orbifold breaking}\label{gt}

The general setup of the `orbifold' breaking described above
is the modding out or restriction of the space of gauge
field configurations by a discrete transformation $K$.
If this discrete group is to be a symmetry of the gauge action
then, in general, it acts as a linear transformation on the Lie
algebra $A^a(x,y) T^a \to A^a(x,K[y]) M^{ab} T^{b}$
that preserves the structure constants, 
$M^{ad} M^{be} f^{def} =  f^{abc} M^{cf}$.
In other words, the action of $K$ on the Lie algebra ${\cal L}(G)$
of the group $G$ must be an automorphism of ${\cal L}(G)$~\cite{Fuchs}.

Such automorphisms of Lie algebras come in two classes, {\em inner}
automorphisms, and {\em outer} automorphisms, the difference
between the two classes being that inner automorphisms can
always be written as a group conjugation $T^a\to g T^a g^{-1}$
for some $g\in G$, while outer automorphisms cannot be so written.
As we will discuss below, the SU(5) `orbifold-GUT' breaking so far
employed in the literature is of the inner-automorphism form.

\subsection{Orbifold breaking by inner automorphisms}\label{inner}

Before describing the possibilities allowed to us
by outer automorphisms, we first discuss the interesting
physics of orbifold actions by $K=Z_n$ inner automorphisms.
While for $K=Z_2$, the matrix $M$ must have
eigenvalues $\pm 1$ as it forms a representation of $K$,  
for more general discrete
actions the matrix $M$ can have complex eigenvalues.

As a simple example consider $C=T^2$ the 2d torus defined by the lattice
(in complex coordinates) $n_1 1 + n_2 \exp(2\pi i/6)$ with $n_1,n_2\in
{\bf Z}$ (cf.~\cite{orbifold}).  In addition to translations through lattice 
vectors, this torus has a ${\bf Z}_3$ discrete rotation symmetry $z
\to \exp(2\pi i/3) z$, by which we can mod out.  This leads to 3 fixed
points at $f_0 =0$, $f_1 = \exp(\pi i/6) /\sqrt{3}$, and
$f_2 = 2 \exp(\pi i/6) /\sqrt{3}$.  Consider the fixed point at the
origin $f_0$: this (like the other fixed points in this case) is left
fixed by the group elements
$F_1 = 1, w, w^2$.  Now suppose that we have an SU(2) gauge theory
on this space (see also the recent analysis of~\cite{kk}), and we define 
the action of the orbifold group on the generators of SU(2) via
\beq
T^a \to\left( \begin{array}{cc} w & 0 \\ 0
& w^2 \end{array} \right) T^a \left( \begin{array}{cc} w^{-1} & 0 \\
0 & w^{-2} \end{array} \right) ,
\label{eq:orbex}
\eeq
where $w$ is a third root of unity, $w^3=1$.  Then the
the subgroup of SU(2) that commutes with this action
is the U(1) generated by $\si^3$.
Thus at the fixed point $f_0$ the gauge symmetry is just
the U(1) left invariant by Eq.~(\ref{eq:orbex}), while away
from the fixed points the full SU(2) is a good symmetry.

As is well known from the string orbifold literature (see
\eg~\cite{ibanez}), the general
structure resulting from such orbifolding is most easily
illuminated by choosing the Cartan-Weyl basis for the generators $T^a$
of the bulk gauge group $G$.  In this basis the generators are organized
into Cartan sub-algebra generators $H_i$, $i=1,\cdots,{\rm rank}(G)$,
and `raising and lowering' generators $E_\al$, $\al=1,\cdots,
({\rm dim}(G)-{\rm rank}(G))$, with
\beq
[H_i, E_\al]=\al_i E_\al\,,
\label{eq:root}
\eeq
where the ${\rm rank}(G)$-dimensional vector $\al_i$ is the root 
associated to $E_\al$. The orbifold action on the gauge and matter 
fields, $A \to g A g^{-1}$ and $\Phi \to g \Phi$, is given by 
a matrix representation $g$ of the action of the $Z_n$ group.

It is always possible to express the action of this discrete
$Z_n$ Abelian group in terms of the Cartan generators as
\beq
g = e^{-2\pi i V \cdot H }\,,
\label{eq:cartan}
\eeq
where this defines the rank$(G)$-dimensional twist vector $V_i$ that
specifies the orbifold action.  (When this twist acts on a field
in representation $r$ of $G$ the $H_i$ are to be considered as in this
representation too.)  Via the use of standard commutator identities
Eqs.~(\ref{eq:root}) and (\ref{eq:cartan}) then imply
\bea
g E_\al g^{-1} &=& \exp\left(-2\pi i \al \cdot V\right) E_\al,\nonumber\\
g H_i g^{-1} &=& H_i ,
\label{eq:znaction}
\eea
so the Cartan-Weyl basis diagonalisies the $Z_n$ action,
and we necessarily have that
$\exp(-2\pi i \al \cdot V) = w$, where $w$
is an $n$-th root of unity.

From Eq.~(\ref{eq:znaction}) we see that gauge bosons corresponding
to Cartan generators are not projected out, since if $T^a=H_i$ is a Cartan
sub-algebra element then the $g$'s commute through $H_i$, and the
transformation acts on $H_i$ as the identity. Thus we
immediately learn that breaking by $Z_n$ inner automorphisms is
{\em rank preserving} (in particular this is true in the $Z_2$ case,
a prototypical example of such inner automorphism action being
precisely the action that reduces SU(5) to SU(3)$\times$SU(2)$\times$U(1)
in the $S^1/(Z_2\times Z_2)$ case). Additionally,
raising and lowering generators with roots $\al$ satisfying
\beq
\al \cdot V = 0 \quad {\rm mod}~{\bf Z}
\eeq
are not projected out.  Thus the problem of determining the unbroken
subgroup is reduced to simple linear algebra. 

In fact there exists a
well-known algorithm for computing the surviving
group under such a $Z_n$ twist.  Consider the {\em extended Dynkin
diagram}\footnote{
See Table~16 of Ref.~\cite{Slansky}, and the
discussions of \cite{Fuchs} and \cite{DMR} where the structure
of the subalgebras of a Lie algebra are discussed in this
language in some detail.
}, 
where in addition to the usual Dynkin
diagram nodes corresponding to the simple roots of the algebra
we add one more node formed from the lowest root $-\al_{\th}$ (where
$\al_\th$ is the highest root, which is not simple).  Then the
regular semi-simple subalgebras of a Lie algebra are almost always found
(the 5 exceptions are discussed in \cite{DMR} and references therein)
by deleting one or more nodes of this extended Dynkin diagram. 
Thus the following simple rule almost always applies: If one
desires to realize the semi-simple subalgebra $H$ with Dynkin
diagram which corresponds to the deletion of nodes $\al_I$ ($I$ runs
over some subset of $\th,1,2,....r={\rm rank}(G)$) one utilizes a
twist $V$ satisfying
\bea
\al_I \cdot V &\neq & 0 \quad {\rm mod}~{\bf Z}, \nonumber \\
\al_i \cdot V &= & 0 \quad {\rm mod}~{\bf Z},~~\forall i\neq I .
\label{eq:subalg}
\eea
In fact these conditions are just the statement that the twist
vector $V$ is a valid weight vector
of the subalgebra $H$ but not of the original algebra $G$.
Moreover, we can write $\al_I \cdot V = k_I/n$, for some integer
$k_I$'s since $V$ has to represent the $Z_n$ twist.  Then gauge
bosons corresponding to such roots $\al_I$ have their KK mode
spectrum lifted by $k_I/nR$, thus eliminating the zero mode.

As a simple example of this consider an SO(10) gauge theory
acted upon by a $Z_4$ twist.  The extended Dynkin diagram of this theory
is shown in the 4th figure of Table 16 of Slansky~\cite{Slansky}. 
If the node indicated by `3' is removed then the
surviving algebra is the Pati-Salam group SU(4)$\times$SU(2)$\times$SU(2).
Since the simple roots of SO(10) are just the rows
of the Cartan matrix of SO(10) (see Tables~6 and 8 of Ref.~\cite{Slansky}),
namely $(2,-1,0,0,0)$, $(-1,2,-1,0,0)$, $(0,-1,2,-1,-1)$, $(0,0,-1,2,0)$,
$(0,0,-1,0,2)$ in the Dynkin basis, while the lowest root is
$\al_\th =(0,-1,0,0,0)$, then the $Z_4$ twist vector
$V=(1/2,0,1/2,1/4,1/4)$ projects out just the third root leaving
the unbroken Pati-Salam group in the zero mode sector.

In the case of $Z_2$, a classification of all breaking patterns is given in 
Table 17 of~\cite{Slansky}. Here we only reproduce the breaking patterns 
for SU(N), SO(N) and E$_6$, since they are most likely to be of physical 
interest.
{\begin{center}
\begin{tabular}{|c|c|c|}
\hline
$G$ & $H$ & restrictions\\
\hline
 SU(p+q) &  SU(p)$\times$SU(q)$\times$U(1)  &  \\ 
 SO(p+q) &  SO(p)$\times$SO(q) & p or q even \\ 
 SO(2n)  & SU(n)$\times$U(1) & \\
 E$_6$ & SU(6)$\times$SU(2) & \\
 E$_6$ & SO(10)$\times$U(1) & \\
\hline
\end{tabular}
\end{center}}
\vspace{3mm}
Table 2. Possible breaking patterns of SU(N), SO(N) and E$_6$ based on 
orbifold action by $Z_2$ {\em inner} automorphisms.

The feature of rank-preservation by inner automorphisms is only
generally true in the case where the orbifold action in field
space is $Z_n$.  For more general actions and, in particular, if this 
action is non-Abelian, it is not possible to write the group element $g$ 
as the exponential of the Cartan subalgebra generators, cf. 
Eq.~(\ref{eq:cartan}), and therefore conjugation by $g$ can act 
non-trivially on some $H_i$, leading to the projection of the 
corresponding gauge fields out of the zero mode spectrum.  If we wish to 
have the gauge symmetry broken on the fixed branes (rather than in the bulk 
as a whole), it is necessary that the orbifold action on the spatial 
coordinates $y$ also be non-Abelian. Such models (which are possible, 
e.g., in higher dimensions~\cite{wit}) can be quite complicated
and highly constrained by anomaly cancellation considerations (as
discussed in Section.~\ref{anomalies}), so despite their potential interest
we now focus attention on a more elementary method of rank reduction.

\subsection{Orbifold breaking by outer automorphisms}\label{outer}

Even if we restrict to Abelian orbifold groups, rank preservation
is {\em not} automatic.  The new possibility of
breaking rank in the Abelian case is realized if
outer automorphisms are employed, and, as we will now show, even
$Z_2$ `orbifold GUT breaking' can reduce the rank. 

Recall that outer automorphisms are structure-constant preserving
linear transformations of the generators which cannot be written
as group conjugations.  For any given Lie algebra there are only
a limited number of possible outer automorphisms, their group structure
corresponding to the symmetries of the Dynkin diagram of the Lie algebra.
To make this concrete, consider the prime
example of an outer automorphism; complex conjugation,
$T^a \to -(T^a)^*$ for all $a\in {\cal L}(G)$,
which of course preserves the structure constants.  For groups with
complex representations this cannot be written as a conjugation
by a group element.  As a simple example consider an SU(n) gauge theory
defined on the orbifold $S^1/Z_2$, with
\bea
T^a A_\mu^a(x,y) &\sim & -(T^a)^* A_\mu^a(x,-y) \cr
T^a A_5^a(x,y) &\sim & (T^a)^* A_5^a(x,-y) .
\eea
Then the gauge fields corresponding to generators in SO(n) but not
SU(n) are even, and have a Kaluza-Klein decomposition containing
zero modes, while the others are odd and possess no zero modes.  Thus
the theory on the brane only respects the gauge symmetry of the SO(n)
subgroup, and the rank is reduced (for $n>2$). In fact in a sense
a U(1) theory provides an even simpler, though somewhat degenerate
example of this.  It is certainly possible in principle to mod out
by the equivalence $A_\mu(x,y) \sim - A_\mu(x,-y)$, $A_5(x,y)
\sim A_5(x,-y)$ which eliminates the U(1) completely from the zero
mode spectrum.  

Of course a requirement for such modding out to make sense is that
the original bulk theory be symmetric under the field space transformation
being employed.  In the case of U(1) including
charged matter the original theory must then have a $q\to -q$
`charge-conjugation' symmetry.  Similarly in the non-Abelian situation
the pure gauge case is trivially consistent as the adjoint representation
is always real, while with matter present we require a $r\leftrightarrow
{\bar r}$ symmetry.  Thus we see that the original bulk theory must
be vector-like, at least with respect to the group we wish to act
on by an outer automorphism.  One may be concerned that this is 
a difficulty if one wants to realize a chiral theory in the zero-mode
sector.  However we are used to the fact that orbifolding can produce
chiral states from an originally vector-like theory, an example being
the chiral $N=1$ theories in 4d resulting from the `$N=2$' minimal SUSY
theory (8 supercharges) in 5d.  The simplest possibility is just
to add chiral matter of the $H$ subgroup theory (in an anomaly-free
representation) on the brane where $G\to H$ via the outer automorphism.

A natural question is if this most simple form of $Z_2$ rank
reduction can work for SO(10), reducing the theory to SU(5) or
SU(3)$\times$SU(2)$\times$U(1) without an additional U(1). 
Unfortunately this does not appear to be possible.  A complete
listing of such $Z_2$ outer automorphism reductions is given in 
Table~2.
{\begin{center}
\begin{tabular}{|c|c|c|}
\hline
$G$ & $H$ & action or restrictions\\
\hline
 U(1) &  1  & $q\to -q$ \\ 
 SU(n) &  SO(n) & $R\to {\bar R}$ \\ 
 SO(p+q) & SO(p)$\times$SO(q) & $R \to {\bar R}$, $p,q$ odd $p+q=4n+2$ \\
 SO(p+q) & SO(p)$\times$SO(q) & $S\to S'$, $p,q$ odd $p+q=4n$\\
 SU(2n) & Sp(n) & $R\to {\bar R}$ \\
 E$_6$ & Sp(4) & $R\to {\bar R}$ \\
 E$_6$ & F$_4$ & $R\to {\bar R}$ \\
\hline
\end{tabular}
\end{center}}
\vspace{3mm}
Table 2. Allowed rank reduction by orbifold action employing
$Z_2$ outer automorphism twist.

\section{Symmetry breaking by boundary conditions}\label{boundary}

As has been discussed in detail in the previous sections, the constructs
resulting from GUT breaking by orbifolding are, in essence, higher
dimensional field theories defined on $R^4\times(C/K)$, where $(C/K)$ is a 
compact manifold with boundaries. Apart from the obvious possibility of having
certain degrees of freedom confined strictly to a boundary, the interesting
structure of these theories is due to the different types of boundary
conditions for the bulk fields. Until now, we have simply accepted that
these boundary conditions are determined by the discrete orbifold
symmetry $K$ (more precisely, by its realization in field space) which
defines $C/K$. The advantage of this approach
is that one simply restricts the space of physical field configurations of
a consistent theory on the basis of a symmetry of the action, thereby
automatically obtaining a new consistent theory.

However, it is not at all obvious that all consistent field theories on
spaces with boundaries can be obtained in this way. Thus, it may be more
general and, in certain cases, more economical\footnote{For
example, to realize the prototypical $S^1/(Z_2\times Z_2')$ orbifold model,
including fields localized on both boundaries, one has to start with a
field theory on an $S^1$ with 4 branes (located at $0$, $\pi/2$, $\pi$ and
$3\pi/2$) the Lagrangians of which are pairwise related by the two $Z_2$
symmetries.}
to start directly with a field theory on $R^4\times M$ (where $M$ is a 
compact manifold with boundary and its possible construction as $C/K$ is 
inessential). This theory is made consistent by an
appropriate choice of boundary conditions. In the present section, we
investigate the possibilities for choosing such boundary conditions and
their implications for the surviving gauge symmetry and particle
spectrum.

\subsection{Consistent boundary conditions for scalar and gauge 
fields}\label{gc}

Let us start with the simple case of a scalar field in a 5d space-time
with 4d boundary located at $y=0$,
\beq
S=\int\limits_{y=0}dy\int d^4x\left(\frac{1}{2}(\partial\phi)^2-V(\phi)
\right)\,.
\eeq
Varying this action, one obtains
\beq
\delta S=-\int d^4x\,(\partial_y\phi)\,\delta\phi\Bigg|_{y=0}-
\int\limits_{y=0}dy\int d^4x\left(\partial^2\phi+V'(\phi)\right)\delta\phi
+\cdots\,,\label{bou}
\eeq
where the dots stand for the contribution from a possible further boundary,
which is of no concern at the moment. We want our theory to be consistently
defined entirely in terms of the bulk field $\phi(x,y)$ with $y>0$. This
will be the case if the boundary contribution in Eq.~(\ref{bou})
vanishes (at least on the level of the classical action; see
Section~\ref{consistency} for a discussion of the quantum case).
Therefore we have the
two obvious possibilities of demanding either $\delta\phi=0$ (Dirichlet)
or $\partial_y\phi=0$ (Neumann) at $y=0$. In the first case, a natural
more special choice is $\phi=0$ at $y=0$.

Generalizing the above to the case of an Abelian gauge theory, one finds
the two analogous possibilities $A_\mu=0$ or $F_{5\mu}=0$ at the boundary.
While the first choice breaks gauge invariance at the boundary, the second
choice leaves the gauge invariance completely intact. This can now be
compared to what is done in the orbifolding case. On the one hand, gauge
symmetry breaking is realized by the parity assignment $A_\mu\to-$ and
$A_5\to+$, which leads to the boundary conditions $A_\mu=0$ and $\partial_y
A_5=0$. It is immediately clear that this is precisely our Dirichlet-type
boundary condition supplemented with the gauge choice $\partial_yA_5=0$
(which can be realized even in the broken theory since the parameter $\chi$
of the gauge transformation $A_M\to A_M+\partial_M\chi\,$, $M=0,1,2,3,5$ 
remains unrestricted
away from the boundary). On the other hand, unbroken gauge symmetry follows
from the assignment $A_\mu\to+$ and $A_5\to-$ and boundary conditions
$\partial_yA_\mu=0$ and $A_5=0$. Again, we see that this is just our
Neumann-type boundary condition in the $A_5=0$ gauge. Thus, at least in
this particularly simple case, the boundary-condition-based approach
describes the same physics.

In the case of a non-Abelian gauge theory with gauge group $G$ one finds,
in complete analogy, that the boundary term in $\delta S$ vanishes if
either $A^A_\mu=0$ or $F^A_{5\mu}=0$ at the boundary ($A_\mu=A_\mu^AT^A$
is the gauge potential and the $T^A$ form an orthonormal basis of the Lie
algebra ${\cal G}$). Let $G$ have a subgroup $H$ with Lie Algebra ${\cal
H}$ so that ${\cal G}={\cal H}\oplus{\cal H}'$. One now has the
phenomenologically interesting option of breaking $G$ to $H$ by demanding
$A_\mu\in{\cal H}$ and $F_{5\mu}\in{\cal H}'$ at the boundary. It is
immediately clear that these conditions are indeed invariant under the full
set of gauge transformations from $H$. Furthermore, let the set of indices
$\{A\}$ consist of $\{a\}$ and $\{\hat{a}\}$ so that the $T^a$ and $T^{
\hat{a}}$ form a basis of ${\cal H}$ and ${\cal H}'$ respectively. Then our
boundary conditions read $F^a_{5\mu}=0$ and $A^{\hat{a}}_\mu=0$, which is
again equivalent to the familiar conditions from orbifolding with an
appropriate gauge choice. Note, however, that one now has vastly more
freedom as far as the symmetry breaking pattern is concerned. While, as
discussed before, only very special subgroups can be obtained by $Z_2$
orbifold breaking, any subgroup $H\subset G$ can be preserved by the boundary
condition breaking described above.

A bulk field $\Phi$ (with components $\Phi_i$) transforming in some
representation of $G$ can be
discussed along the same lines. The boundary term in $\delta S$ vanishes if,
for each $i$, either $\Phi_i$ or $(D_y\Phi)_i$ (where $D$ is the covariant
derivative) vanishes at the boundary. More abstractly, if $\Phi$ takes
values in the vector space $V=V_1\oplus V_2$, than we can demand
$\Phi\in V_1$ and $D_5\Phi\in V_2$ at the boundary. Given that $G$ is
broken to $H$ in the gauge sector, no further symmetry breaking will be
introduced by this choice if $H$ is represented on $V_1$ and $V_2$
separately. Again, this is much more general than what is possible with the
$Z_2$ parity matrix $P$ familiar from orbifolding. For example, one can choose
$V_1=0$ ($V_2=0$) so that no (all) components of $\Phi$ have KK zero modes.

\subsection{Are more restrictive boundary conditions possible?}
\label{generalbc}

In the previous two subsections we have discussed the imposition
of either Neumann or Dirichlet boundary conditions on the
bulk fields at the location of the brane.  The most interesting case is 
the breaking of a non-Abelian gauge group $G\to H$ by $F^a_{5\mu}=0$ 
and $A^{\hat{a}}_\mu=0$.  In the bulk, the allowed gauge
transformations are of the form $U=\exp(i \sum_{a}\xi^a(x,y) T^a + i
\sum_{\hat{a}} \xi^{\hat{a}}(x,y) T^{\hat{a}} )$, with both the gauge
transformation parameters $\xi^a(x,y)$ and $\xi^{\hat{a}}(x,y)$
non-vanishing.  Only at the brane are the $\xi^{\hat{a}}(x,\ell) = 0$,
and purely the SM gauge symmetry is defined.  However it is misleading
to say in this case that there is no remnant of the bulk gauge
symmetry $G$ at the brane.  The reason is that the $y$-derivatives
of the gauge field in the broken directions is non-zero
$\p_y A^{\hat{a}}_\mu\neq 0$, and thus in general there can
brane-localized interactions involving such a combination.
Therefore, an interesting question is if it is possible to impose
more restrictive boundary conditions that eliminate simultaneously
both $A^{\hat{a}}_\mu$ and $\p_y A^{\hat{a}}_\mu$.\footnote{We
thank Lawrence Hall for raising this question and for discussions.}

To understand the problems of such a setup it is sufficient to address the 
case of a bulk scalar field. Let us therefore impose both $\phi=0$ and 
$\p_y\phi=0$ at the boundary. According to the discussion in Sect.~\ref{gc}, 
this is certainly a self-consistent boundary condition at the level of the 
classical action. However, by doing so one excludes completely the 
existence of the familiar KK excitations. To see this decompose the field 
in to 4d momentum eigenstates, so that $\p_\mu \p^\mu\phi=-k^2\phi$.  Then 
the 5d field equation becomes (at the linearized level, suitable for 
building perturbation theory)
\beq
-(\p_y^2-k^2) \phi = m^2\phi
\eeq
with $m^2 = \p^2 V(\phi)/\p\phi^2 |_{\phi=0}$. With the above boundary 
condition, the only solution is $\phi(x,y)\equiv 0$. This implies that 
the space of 5d field configurations can not be described in terms of a 
superposition of eigenfunctions of the 4d momentum operator $\hat{P}_\mu$ 
(with eigenvalues $k_\mu$.) Since usually this is the starting point of 
the quantization procedure, it is not obvious how to quantize the theory 
in the sense of conventional weakly coupled quantum field theory. Note 
that this problem is not improved in the more general situation with 
$\phi=\phi_0\neq 0$ at the boundary (let alone the fact that such a 
boundary condition is unnatural in the physically interesting case of a 
gauge potential $A_\mu$). 

To summarize, the more restrictive boundary conditions described above 
appear to work on a classical level, but not in any straightforward way
in the quantum theory.

\subsection{Boundary condition breaking from boundary VEVs}
\label{boundaryvev}

A simple field-theoretic realization of the above symmetry breaking by
boundary conditions is obtained if a gauged boundary scalar $\Phi$ is
included in the action,
\beq
\Delta S=\int\limits_{y=0}dy\int d^4x\,\delta(y)\,|D\Phi|^2\,,
\eeq
and $\Phi$ acquires a vacuum expectation value (VEV)~\cite{NSW}. Let us again
start with the Abelian case and impose the gauge symmetry preserving
boundary condition $F_{5\mu}=0$. Assume that the complex scalar $\Phi$
acquires the VEV $\Phi=v/\sqrt{2}$. Then, in the $A_5=0$ gauge, a KK mode
of the field $A_\mu$ with 4d momentum $k$ has to solve the differential 
equation
\beq
\left[-\frac{1}{g_5^2}\left(\partial_y^2+k^2\right)+\delta(y)v^2\right]\,
A_\mu=0\,.
\eeq
Integrating this equation from 0 to an infinitesimal positive constant
$\epsilon$ and making use of the boundary condition $\partial_y A_\mu=0$,
one obtains
\beq
\partial_yA_\mu\Big|_{y=\epsilon}=g_5^2v^2A_\mu\,.
\eeq
Thus, in the limit $v^2\to\infty$, the field $A_\mu$ is driven to zero at
the boundary. At the same time, the original boundary condition
$\partial_y A_\mu=0$ is relaxed at infinitesimal distance from the
brane. As a result, we see that a large VEV of a gauged boundary scalar
realizes precisely the gauge symmetry breaking boundary condition discussed
abstractly in Sect.~\ref{gc}.

The non-Abelian case can be discussed in complete analogy. The main
difference is that only the fields $A_\mu^{\hat{a}}$ (where $T^{\hat{a}}$
are those generators which act non-trivially on the VEV of $\Phi$) are
forced to zero by a diverging VEV. Thus, precisely as discussed in
Sect.~\ref{gc}, any breaking pattern can be realized if boundary scalars
in arbitrary representations of $G$ can acquire VEVs. For example, SO(10)
can be broken to SU(5) by a large vacuum expectation value of a brane
localized scalar field in the ${\bf 16}$ of $SO(10)$. This is physically 
different from either inner or outer $Z_2$ orbifold breaking.

\section{The relation between orbifold breaking and Wilson line breaking
of gauge symmetries}\label{wilson}

In this section, we discuss the relation between orbifold breaking of gauge 
symmetries and the familiar Wilson line breaking mechanism (also known as 
flux breaking or the Hosotani mechanism)~\cite{hos,wit,gsw}. This mechanism 
works for a gauge field theory on a space $R^4\times(C/K)$, where $K$ acts 
freely on $C$. There exist two equivalent descriptions, depending on 
whether one considers fields defined on the covering space $R^4\times C$ or 
on the true physical space $R^4\times(C/K)$ (see~\cite{dm} for a 
particularly clear discussion of this issue).

In the first definition, one requires for consistency that fields $\Phi$ 
transforming under the gauge group $G$ are identical up to an 
$x$-independent gauge transformation when evaluated at two points of $C$
related by a transformation $k\in K$:
\beq
\Phi_i(x,k[y])=R_{ij}(k)\Phi_j(x,y)\,.\label{trf}
\eeq
Here $x\in R^4$, $y\in C$, and the map $k\to R(k)$ has to be respect the 
group property: $R(k\cdot k')=R(k)\cdot R(k')$. In particular, 
Eq.~(\ref{trf}) holds for the gauge potential itself, in which case 
$R(k)$ is a matrix in the adjoint representation of $G$. In this approach, 
it is obvious that orbifold breaking is obtained from Wilson line breaking 
by simply giving up the requirement of a {\it free} action of $K$ on $C$. 
Apart from that, the discussion of Sect.~\ref{orbb} applies to both 
orbifold and Wilson line breaking. If the map $k\to R(k)$ is derived from 
a group homomorphism $K\to G$ (cf. the inner automorphism case of 
Sect.~\ref{gt}), then the symmetry is reduced to those elements of $G$ that 
commute with all elements of $K$. 

We observe that, on a technical level, orbifold breaking has one important 
new feature when compared to Wilson line breaking. Since it is assumed that 
the gauge potential is continuous~\footnote{
After the first version of this paper appeared, orbifold theories with
discontinuities of certain fields at the fixed points have been discussed 
in~\cite{bfz}. Such structures represent a modification of the original 
orbifolding idea where all fields are smooth because they belong to a 
subset of the field space on a smooth manifold.}, 
gauge fields $A_\mu^{\hat{a}}$ on which 
$K$ acts non-trivially are forced to actually vanish at the orbifold fixed 
point. This has no analogue in Wilson line breaking because of the {\it 
free} action of $K$ on $C$. 

In the second definition of Wilson line breaking, fields are defined on 
$R^4\times(C/K)$ in the presence of a background field $B_M$ with vanishing 
field strength. By definition, Wilson loops of $B_M$ corresponding 
to certain non-contractible closed paths in $C/K$ take certain non-trivial 
fixed values (which explains the name Wilson line breaking). The relation to 
the first definition follows from the observation that non-contractible 
loops in $C/K$ can be lifted to paths connecting points $y$ and $k[y]$ in 
$C$, and the corresponding Wilson lines provide the desired group 
homomorphism $K\to G$. 

Starting with the first definition of Wilson line breaking, one can go to 
the second definition by performing a gauge transformation on $C$ that 
undoes the relative rotation of fields at $y$ and $k[y]$. The vacuum on 
$C$ (where the gauge potential vanishes) is thereby transformed into the 
background field $B_M$ discussed above. More explicitly, let this gauge 
transformation be defined by the function $U(y)$. Then 
\beq
A_M(x,y)\to U(y)[A_M(x,y)-i\partial_M]U^{-1}(y)=A_M'(x,y)+B_M(y)\,,
\eeq
where $B_M$ is the background field and $A_M'$ is the new gauge field, 
satisfying the condition $A_M'(y)=A_M'(k[y])$. Following this line of 
thinking, it is particularly easy to see what the orbifold analogue of the 
second definition of Wilson line breaking looks like. We will now describe 
this procedure in the case of the simple model of SU(5) breaking on 
$S^1/(Z_2\times Z_2')$ of Sect.~\ref{ex}. 

After modding out of the first, SU(5)-preserving $Z_2$, our theory is 
defined on the interval $y'\in[-\ell,\ell]$. Thus, $C=[-\ell,\ell]$ is the 
covering space on which $K=Z_2'$ acts, and $C/K=[0,\ell]$. According to the 
first definition of Wilson line breaking, we demand that $\Phi_i(x,-y')=
R_{ij}(-1_{Z_2'})\Phi_j(x,y')$, where $R(-1_{Z_2'})=$ diag$(1,1,1,-1,-1)=
-P'$ in the fundamental representation. In particular, this has to hold 
for the gauge potential\footnote{
Note 
that, viewing the gauge potential as a Lie-algebra valued form, 
the extra minus sign of the 5th component $A_5$ becomes a trivial 
consequence of the reflection of space induced by $-1_{Z_2'}$.
}.
Now we can perform a gauge rotation with the matrix $-P'\in$ SU(5) on 
$[-\ell,0]\subset[-\ell,\ell]$. This rotation is generated by an element 
of the Cartan subalgebra of the Lie algebra of SU(5): $-P'=$ exp$(iT)$ with 
$T=\pi$ diag$(2,2,2,-3,-3)$. As a result, we now have a theory where 
$\Phi_i(x,-y')=\Phi_j(x,y')$. The price we pay for this is that we have to 
work in the non-trivial background $B_M=\delta_{5M}\delta(y')T$. In other 
words, we work in a background with non-trivial Wilson loop
\beq
W=\exp\left(i\int_{-y'}^{y'}dy''B_5(y'')\right)=-P'\,.
\eeq
Since, by definition, we are now discussing a theory where the gauge 
potential is symmetric under $Z_2'$ (i.e., $B_5(-y')=-B_5(y')$ for $y'\neq 
0$), this Wilson loop gets a contribution only from the orbifold fixed 
point. It is also, as required, invariant under gauge transformations on 
$C/K\equiv[0,\ell]$. This becomes obvious if one observes that, when 
working on the covering space $C\equiv[-\ell,\ell]$, such gauge 
transformations rotate fields on $(0,\ell]$ and $[-\ell,0)$ in the 
same way.

Now consider the classical action for a perturbation $A_\mu$ (recall that 
now $A_\mu(y')=A_\mu(-y')$) in this background. Since 
\beq
F_{5\mu}^A=\partial_5 A^A_\mu-\partial_\mu A^A_5+f^{ABC}(B^B_5+A^B_5)
A^C_\mu\,,
\eeq
the term $F_{5\mu}F^{5\mu}$ in the Lagrangian induces a boundary mass for 
fields $A^{\hat{a}}_\mu$ (where $T^{\hat{a}}$ are those generators which do 
not commute with $T$.) More specifically, the lagrangian for 
$A^{\hat{a}}_\mu$ up to quadratic order reads
\beq
{\cal L}=-\frac{1}{4g^2}\sum_{\hat a}\left[\partial_\mu A_\nu^{\hat{a}}-
\partial_\nu A_\mu^{\hat{a}}\right]^2-\frac{1}{4g^2}\sum_{\hat{a}}\left[
\partial_5 A_\mu^{\hat{a}}+\sum_{b,\hat{c}}f^{\hat{a}b\hat{c}}B_5^b
A_\mu^{\hat{c}}\right]^2\,.\label{qlag}
\eeq
We see that the term proportional to $B_5^2$ corresponds to a mass term for 
the field $A^{\hat{a}}_\mu$. This mass term is localized at $y'=0$ and
proportional to the square of a $\delta$ function. We conclude that this 
infinite mass term will force $A^{\hat{a}}_\mu$ to vanish at the boundary 
$y'=0$. To be more precise, one can replace the $\delta$ function by 
a function $\delta_\epsilon(y')$, defined to be $1/\epsilon$ for $y'\in
(-\epsilon/2,\epsilon/2)$ and zero otherwise, and take the limit 
$\epsilon\to 0$ at the end. Now consider field configurations 
$A^{\hat{a}}_{\mu(\epsilon)}$ which are finite together with their first 
derivatives in the limit $\epsilon\to 0$. It is clear that one has to 
require $\lim_{\epsilon\to 0}A^{\hat{a}}_{\mu(\epsilon)}(y'=0)=0$ to obtain 
a finite action in this limit. Thus, as in the orbifolding definition of the 
model, the fields $A^{\hat{a}}_\mu$ corresponding to broken generators of 
the gauge group $G$ are forced to vanish at the boundary.~\footnote{
One could reach the same conclusion by analyzing the equation of motion 
for $A^{\hat{a}}_\mu$ in the regularized-$\delta$-function background and 
demanding finiteness of the field and its first derivatives in the limit 
$\epsilon\to 0$.}

Note that the infinite mass term in Eq.~(\ref{qlag}) can be understood as 
coming from the kinetic term $|D_\mu H|^2$ of an adjoint Higgs field $H$ 
at the boundary which develops an infinite expectation value in $T$ 
direction. Clearly, on a technical level, the background field $B_5$ plays 
the role of this Higgs field. Thus, similarly to the situation on smooth 
manifolds~\cite{wit}, there exists an analogy between Wilson line breaking 
and breaking by an adjoint Higgs VEV.

In this section, we have so far assumed that $K$ acts on all fields by 
$x$-independent gauge transformations (cf.~Eq.~(\ref{trf})). However, in 
the particularly interesting orbifolding example of Sect.~\ref{ex} this is
obviously not the case. In fact, it is precisely the choice of the matrix 
$P'$ (where $-P'\in$ SU(5)) for the transformation of the Higgs which so 
elegantly solves the doublet-triplet splitting problem in Kawamura's 
original proposal~\cite{kaw1}. We want to interpret this construction as a 
particular example of Wilson line breaking. To achieve this, let us 
generalize what we mean by Wilson line breaking by allowing matrices 
$R_{ij}(k)$ in Eq.~(\ref{trf}) corresponding to all the symmetries of the 
theory (and not just gauge transformations). 
All that matters is that the group multiplication in $K$ is respected and 
that the Lagrangian is left invariant. From this perspective, the orbifold 
model of Sect.~\ref{ex}, now including the Higgs field, is conceptually the 
same as Wilson line breaking (more generally understood), but with a 
non-free action of $K$. In fact, one could take the obvious next step and 
address the doublet-triplet splitting problem in an SU(5) GUT with 
conventional Wilson line breaking on an $S^1$ by introducing an additional 
phase $-1$ in the boundary conditions of the Higgs. 

Such additional phases are, of course, familiar in the case of fermions, 
where they do not affect observable quantities since these are quadratic 
in fermion fields. However, we see no conceptual problem with introducing 
them for scalar fields as well. One may certainly worry whether it is 
still valid to consider $R^4\times S^1$ as the true physical space, given 
that the (observable) sign of a scalar field is not unambiguously defined. 
This may, however, not be necessary for the theory to be consistent.

\section{(Quantum) Consistency of field theoretic
orbifolds}\label{consistency}

In string theory a well-understood issue is that of the consistency
of string propagation on singular orbifold spaces.  In the effective
field theory context in which we are here working this issue needs
to be re-examined.  First, though, we should quickly
address why we choose to work in the effective theory 
framework at all.  

If one is interested in investigating
the phenomenological consequences of a particular field content and
symmetry structure, the effective theory approach is in many
ways superior: it allows an efficient survey of the possibilities
unencumbered by unnecessary restrictions that would be inferred from any
single UV completion.\footnote{Of course the
construction of a specific UV theory can give us new relations
between couplings in the effective theory which can be of great
interest, just like QCD gives relations between couplings in the
low-energy effective chiral Lagrangian theory of pions.}
In particular, if one finds that a phenomenologically
desirable structure is impossible to realize in the effective-field-theory
context, then this approach assures one that it is also impossible to
realize in any UV completed theory. 
Moreover, we claim that all necessary consistency conditions that
the low-energy theory must satisfy {\em can already be seen at low-energy}.
An outstanding example of such a constraint is the necessity of anomaly
cancellation in the low-energy effective theory.    

We emphasize that this philosophy is nothing but the usual one
for effective field theories,
and which has had great success since at least the era of the Fermi
model of the weak interactions, and the Ising and Heisenberg models
of magnetic materials. 

\subsection{Unitarity, self-adjointness, and boundary
conditions}\label{unitarity}

We now turn to the specific question of the field theoretic consistency
of orbifold models.  Most elementary is the requirement that the
truncation enacted by the orbifold projection on the space of KK modes
be consistent with the interactions of the theory.  If not then, for example,
it would be possible to scatter two allowed zero modes producing one or more
disallowed zero modes, and the pole and cut structure of $S$-matrix elements
would not correspond to physical particle states.
This requirement is automatically met if the
orbifold action mods out by a symmetry of the parent theory.  
In the boundary condition approach of Section~\ref{boundaryvev}
it is also guaranteed by, for example, the consistent realization
of the necessary boundary VEVs via a boundary quantum field theory. 

At the quantum level there seem to be two different
types of consistency questions:  First one may be concerned about
loss of unitarity in propagation on such singular spaces.  Certainly
large enough co-dimension orbifold fixed points are locations of
curvature singularities, and are thus not in the usual sense geodesically
complete.  For example in co-dimension two (\eg, the 6d model
$R^4\times T^2/Z_3$ discussed in Section~\ref{inner}) the
fixed point is a conical singularity.  Thus in the low energy
effective Lagrangian there are naively $\de$-function
terms in the curvature, which can appear in the effective action
for the light fields, leading
to ambiguous time evolution.\footnote{Consistent with our effective field
theory viewpoint, these $\de$-functions
should really be interpreted as some distribution with characteristic
length scale $1/M_*$, where $M_*$ is our gravitational cutoff.}
The real issue is, however, possible violation of the 
conservation of energy and momentum (or other conserved quantum numbers)
at the singularity in the second-quantized theory.  To assess if
this happens we must
ask if it is possible to apply boundary conditions to the fields at
the singularity so as to ensure that operators such as the 
Hamiltonian or angular momentum are self-adjoint. More precisely
in the space of $L^2(R)$-integrable functions $\phi$ we require that
it is possible to find boundary conditions such that
\beq
\int_0^r \sqrt{g_{ij}} \phi^* (H \phi) =
\int_0^r \sqrt{g_{ij}} (H^\dagger\phi^*) \phi,
\eeq
where $r$ is here a radial coordinate running from the position
of the singularity, taken to be at $r=0$.  
Mathematically speaking, this is the problem
of finding the so-called self-adjoint extensions of an
operator~\cite{sing,AMRW,dSGJ,CK}.  Such a self-adjoint
extension is typically allowed when the singularity is not too severe. 
Specifically, as demonstrated in Ref.\cite{dSGJ} for the case
of conical gravitational singularities,
and in Ref.~\cite{AMRW} in the case of gauge singularities of
Aharonov-Bohm type (namely gauge configurations with non-zero Wilson lines,
and $F=0$ except at co-dimension 2 singular points -- which we know
by the discussion in Section~\ref{wilson} do occur at the
fixed points when we act on the fields by gauge twists), the
self-adjoint extension does exist.
This ensures that there is no problem with unitarity loss in these
theories with conical-type singularities.\footnote{Quite
often the self-adjoint extension is not unique, but is instead described
by a multi-parameter family.  This just corresponds
to the fact that low-energy unitarity is not strong enough
to fix the boundary conditions uniquely, and these additional
parameters simply correspond to the fact that
a given IR theory may have different UV completions.}
Actually, to be precise the effective orbifold theory is unitary at
{\em low-energy}, below the cutoff.  From an effective theory
viewpoint this is all we need care about: high-energy unitarity
is not in the domain of concerns for an effective theory.  For example,
$S$-matrix elements derived from the chiral Lagrangian describing pion
interactions have, at energies comparable to the cutoff, $E\sim 1\gev$,
unphysical poles and cuts.  This does not invalidate the
theory as a good description below the cutoff.

We now turn to the second consistency issue, that of anomaly
cancellation.

\subsection{Anomaly constraints}\label{anomalies}

The absence of anomalies in the low-energy theory is a more restrictive
and interesting requirement.  It is important to understand that
in orbifold models there are in principle two types of anomalies:
4d anomalies~\cite{anomalies} intrinsic to the fixed points, and
higher-dimensional
anomalies intrinsic to the bulk.  The basic reason for there being two
classes of anomaly cancellation requirements is that the orbifold action
can introduce new chiral fields localized at the fixed points (3-branes),
in addition to the higher-dimensional chiral fields that may
already propagate in the bulk.  Alternatively,
in the boundary condition approach the boundary quantum field theory
realizing the boundary conditions can (must) be chiral, and can in principle
introduce new anomalies.  It is necessary for the low energy consistency
of the theory that {\em both} the fixed-point and the bulk anomalies
be canceled.  For the 4d fixed point anomalies it is sufficient
to check anomaly cancellation only for the massless zero mode
spectrum, since it is only massless fields that cannot be regulated in 
a gauge-invariant way (say by Pauli-Villars) that can lead to an anomaly.
Such 4d anomaly cancellation conditions are identical to those usually
imposed on the SM, and since we want to realize the
spectrum of the SM or MSSM this is not a particular difficulty.
(See the discussion of Ref.~\cite{AHCG}, where, however, considerations
were limited to the anomalies associated to the fixed points.)
However, in space-time dimension $d>5$ the cancellation of the bulk
local anomalies turns out to be a very restrictive requirement. 

To motivate this
let us consider the interesting example of an SO(10) theory (which for
simplicity we choose to be non-supersymmetric) defined on a 6d bulk.
Let us suppose that this
theory has as its left-handed fermion content a single ${\bf 16}$ of SO(10).
If this theory were in 4d then it would be anomaly free.  However, in 6
dimensions the anomalous triangle diagram with 3 gauge currents
is replaced by a box diagram with 4 gauge currents, and, unfortunately,
a single ${\bf 16}$ of SO(10) has a non-zero quartic anomaly as  
the totally symmetric 4th order invariant
${\rm Tr}_S(T^a T^b T^c T^d) = A(\de^{ab}\de^{cd} +\de^{ac}\de^{bd}
+\de^{ad}\de^{bc})$ of SO(10) in the spinor representation is non-zero
(we give its precise value in Eq.~(\ref{eq:sontraces})).

Thus, as an example of the general situation in $d\ge 6$,
let us consider some aspects of the anomaly structure of 6d
theories in greater
detail.  Unlike the Lorentz group in odd dimensions (5d for
example) where chirality is not defined, the 6d Lorentz group
has chiral representations.  In fact there are three types of
fields which contribute to anomalies: chiral spin 1/2 fermions,
chiral spin 3/2 fermions, and (anti) self-dual 3-forms.  
As discussed in Refs.~\cite{AGG} (see also
\cite{SW} for useful trace relations) the total anomaly can be deduced
via so-called descent equations from a formal eight-form
polynomial $I(F,R)$ built out of the Yang-Mills 2-form field strength
$F$ and the 2-form Riemann tensor $R_{\mu\nu}$.\footnote{Traces are
taken over the SO(5,1) indices $a,b$ of $(R_{\mu\nu})_a^b$.}
The 8-form anomaly polynomial is the sum
of the contributions from the various chiral fields in the theory.

It turns out that in the case of pure gauge and mixed gauge-gravity
anomalies only chiral spin 1/2 fermions contribute, leading to
\bea
I_{\rm gauge}^{(1/2)}(F) &=&
-\frac{1}{4! (2\pi)^3 }\tr_r F^4 ,
\nonumber\\
I_{\rm mixed}^{(1/2)}(R,F) &=& \frac{1}{4}
\frac{1}{4! (2\pi)^3} \tr R^2 \tr_r F^2 ,
\label{eq:gaugeanom6}
\eea
where $r$ is the gauge representation of the fermions, and
wedge products are assumed.        
On the other hand there are three sources of the purely gravitational
anomalies arising from chiral spin 1/2, chiral spin 3/2, 
and self-dual 3-form fields.  Their contributions are  
\bea
I_{\rm grav}^{(1/2)}(R) &=& 
-\frac{1}{4! (2\pi)^3}\left(\frac{1}{240} \tr R^4
+\frac{1}{192} (\tr R^2)^2\right) ,
\nonumber\\
I_{\rm grav}^{(3/2)}(R) &=& 
-\frac{1}{4! (2\pi)^3}\left(\frac{49}{48} \tr R^4
-\frac{43}{192} (\tr R^2)^2\right) ,
\nonumber\\
I_{\rm grav}^{({\rm 3-form})}(R) &=&
-\frac{1}{4! (2\pi)^3}\left(\frac{7}{60} {\rm tr}R^4
-\frac{1}{24} (\tr R^2)^2\right).
\label{eq:gravanom6}
\eea
For bulk anomaly cancellation we need the total anomaly
polynomial from the sum over all bulk fields to vanish. 

To make use of these fomulae for anomaly cancellation we need
the relationship between the anomaly contributions from fields
in different SO(10) representations, or, equivalently, the 
relationship between traces taken in different representations.
For SO(n) theories we have for example (denoting the fundamental $F$,
adjoint $A$, and spinor $S$):
\bea
\tr_{A} F^2 &=& (n-2) \tr_F F^2\nonumber \\
\tr_{A} F^4 &=& (n-8) \tr_F F^4 + 3(\tr_F F^2)^2\nonumber \\
\tr_S F^2 &=& 2^{(n-8)/2} \tr_F F^2 \nonumber\\
\tr_S F^4 &=& -2^{(n-10)/2} \tr_F F^4 + 3( 2^{(n-14)/2})(\tr_F F^2)^2 .
\label{eq:sontraces}
\eea
(For completeness the SU(n) case has, for example,
$\tr_{A} F^2 =  2n \tr_F F^2$ and 
$\tr_{A} F^4 =  2n \tr_F F^4 + 6(\tr_F F^2)^2$ 
except in the cases $n=2,3$ where there is no independent
4th-order invariant and $\tr_{A}F^4 = (\tr F^2)^2/2$.)

Thus from Eqs.~(\ref{eq:gaugeanom6}) and (\ref{eq:sontraces}) we see that
a single ${\bf 16}$ of SO(10) has a gauge anomaly.  Let us focus on the
leading term $-\tr_F F^4$ in the expansion of $\tr_S F^4$.  To cancel this
requires either a rhd mirror ${\bf 16}$ (so the theory is non-chiral), or a 
lhd spin 1/2 field in the ${\bf 10}$ of SO(10).  Actually because of the
second term proportional to $(\tr_F F^2)^2$ in the expansion of the spinor
trace of $F^4$ this is not sufficient to cancel the entire anomaly.
Such sub-leading factorized contributions to the anomaly polynomial
correspond to what are known as reducible anomalies, in distinction
to the leading so-called irreducible anomalies.   The reducible
anomalies, which are a new feature relative to the familiar 4d case,
do not necessarily have to cancel by summing over the chiral
matter spectrum.  Rather, the
Green-Schwarz mechanism~\cite{GS} utilizing the interactions and
exchange of antisymmetric tensor fields can apply.  Non-trivial
string-theoretic orbifolds always invoke this mechanism.  Thus we expect
that the introduction of antisymmetric tensor fields will be necessary
to render the effective field-theory orbifold anomaly-free, and these
fields lead to axion-like degrees of freedom
in the low-energy 4d theory.  Although it is quite interesting to
discuss in detail these extra fields and their consequences,
we leave this for a future publication devoted to the
phenomenology of generalized orbifold GUT models.  Here we just note      
that, independent of the possibility of the GS mechanism, it is a necessary
condition for anomaly freedom that the chiral field content of the
theory leads to a cancellation of the irreducible anomalies.  This
alone is a highly restrictive requirement.

In fact the minimally supersymmetric 6d theories
($(1,0)$-SUSY, with 8 real supercharges) of most interest are 
further constrained since the $(1,0)$ SUSY algebra requires
the gravitino and gauginos to have opposite chirality from the
matter fermions which must all share the same
chirality~\cite{sagnotti}.  Taking a Panglossian viewpoint it is
possible to dream that these
severe constraints in the 6d (and higher) case can be used to derive
or restrict the number of generations, along the lines of the discussion
of Ref.~\cite{DP}.

\section{Conclusions}

In this paper we have performed a detailed analysis of the structure
of orbifold gauge theories, paying particular attention to the 
possibilities for gauge symmetry breaking.  To set the stage for our
discussion we studied in Section~2.1 the field-theoretic orbifolding 
procedure. The meaning of `modding out' a theory by a discrete
symmetry group acting simultaneously in physical and
field space was discussed, while in Section~6 we studied issues
connected to the quantum-mechanical consistency of this procedure
in the effective field theory context.  These included the requirements
for defining a sensible theory on singular orbifold spaces, and
especially the stringent anomaly cancellation conditions that apply
to theories in $d>5$ dimensions.

The group theoretic structure of orbifold breaking of gauge symmetries
was analysed in detail in Section~3.  Different breaking patterns
emerge if the discrete orbifolding symmetry is realized in field
space as an inner (see Sect.~3.1) or outer (see Sect.~3.2) automorphism
of the Lie algebra.  A complete classification was given of the simple
but physically very interesting case of $Z_2$ orbifold actions, which
includes the popular $S^1/(Z_2\times Z'_2)$ models.  We showed that rank
reduction is possible, and discussed how this is achieved in the $Z_2$
case by the use of outer automorphisms.  Some aspects of larger
orbifold-GUT theories based on the groups SO(10) and E$_6$ were
noted. (We emphasize that for more general non-Abelian orbifold
actions rank reduction is generic.)

In Section~4, we outlined an alternative and in principle much more 
general approach, which starts from a theory defined on a space with 
boundaries and breaks the gauge symmetry by consistently chosen 
boundary conditions for the gauge potential.  One realization of this 
is by expectation values of boundary fields, as discussed in Section~4.3.
For example, the physically important breaking of SO(10) to SU(5), 
inaccessible to the $Z_2$ orbifolding procedure, is straightforwardly 
realized by a boundary scalar in the {\bf 16} of SO(10).  In Section~5
we turn to the close relation of orbifold breaking (as so far studied
in the context of the $S^1/(Z_2\times Z'_2)$ model) and the more
traditional Wilson line breaking of gauge symmetries.  We argued that,
in the orbifolding case, one can equivalently think of a background gauge 
field restricted to the singularity (with non-trivial Wilson loop)
as enforcing the orbifold boundary conditions for the gauge potential.

Overall, it is clear that GUT theories constructed by field-theoretic
orbifolding have a rich structure with exciting new possibilities for
phenomenology.

\vspace*{1cm}

\noindent
{\bf Acknowledgments}:
We are very grateful to Guido Altarelli, Wilfried Buchm\"uller, Lawrence 
Hall, Takemichi Okui, Erich Poppitz, Riccardo Rattazzi and Claudio
Scrucca for helpful conversations.
JMR thanks the members of the Theory Group, LBNL, for their kind 
hospitality during the completion of this work.

\end{document}